\documentclass{article}

\usepackage{arxiv}
\usepackage[version=4]{mhchem}
\usepackage{cleveref}
\usepackage{booktabs}
\usepackage{authblk}

\newcommand{\sigon}{\ensuremath{\sigma_\mathrm{on}}}
\newcommand{\sigoff}{\ensuremath{\sigma_\mathrm{off}}}
\newcommand{\ENF}{\ensuremath{E_\mathrm{NF}}}
\newcommand{\EBG}{\ensuremath{E_\mathrm{BG}}}

\begin{document}

\pagestyle{fancy}
\title{Time-resolved SNOM via phase-domain sampling}
\date{April 2, 2026}

\author[1]{Philipp Schwendke}
\author[1,2]{Julia St\"{a}hler}
\author[1]{Samuel Palato\thanks{samuel.palato@hu-berlin.de}}
\affil[1]{Institut f\"{u}r Chemie, Humboldt-Universit\"{a}t zu Berlin, Berlin, Germany}
\affil[2]{Abteilung Physikalische Chemie, Fritz-Haber-Institut der Max-Planck-Gesellschaft, Berlin, Germany}

\maketitle

\begin{abstract}
Time-resolved scanning near-field optical microscopy (tr-SNOM) enables the measurement of the dynamic optical response of functional surfaces beyond the diffraction limit.
Experimental challenges are imposed both by the use of a pulsed light source, and by the need for interferometric signal modulation to isolate the near-field contribution.
We present a novel experimental approach to retrieve the tr-SNOM signal using a 200\,kHz laser system and pseudo-heterodyne modulation.
We circumvent the Nyquist limit for spectral demodulation by sampling modulation phases, pump intensity and SNOM signal for every laser shot.
A general time-resolved SNOM signal is derived, independent of detection scheme or physical assumptions about the near-field enhancement, and is successfully measured and isolated on \ce{WS2} monolayer and multilayer regions.
We confirm localization by signal-distance curves, spatial confinement at material boundaries, and by identifying signal contributions at individual modulation harmonics.
Disentangling the dynamic contributions enables us to extract the dynamic dielectric function of the sample.
Showing the capability of phase-domain sampling paves the way to integration of more diverse and specialized light sources, growing the potential of optical ultrafast near-field measurements.
\end{abstract}
\keywords{SNOM, time-resolved, pseudo-heterodyne, phase-domain sampling, \ce{WS2}}

\section{Introduction}
The unique potentials of functional materials, such as 2D materials and their heterostructures, emerge from their electronic structure on the nanoscale.
Scanning probe microscopy is uniquely suited to investigate a large range of material properties, offering simultaneous topographic, optical, mechanic, and electric information\cite{Bar2000,Munz2023}, in a \emph{lab on a tip} approach\cite{Shiotari2025}.
Light-matter coupling is especially interesting on length scales comparable to or below the wavelength of the light, with direct technological applications\cite{Ansari2020,Rothe2020}, but also unique physics, e.g.\ nano-resonators\cite{Verre2019} and polaritons\cite{Wiecha2019}.
While scanning near-field optical microscopy\cite{Hillenbrand2025} (SNOM) pushes the spatial resolution beyond the diffraction limit, ultrafast measurements offer insights into functional processes via signal kinetics.
In a combination, time-resolved SNOM (tr-SNOM) has been used to characterize carrier dynamics in semiconductors\cite{Eisele2014,Pizzuto2021}, follow the insulator-to-metal transition in \ce{VO2}\cite{Doenges2016,Sternbach2021,Nishikawa2023}, and describe exciton dynamics in \ce{WS2}\cite{Mrejen2019,Plankl2021,Wang2024}.

However, ultrafast measurements require a pulsed light source, limiting the sampling frequency by its repetition rate.
In order to resolve higher orders in tapping modulation\cite{Knoll2000}, tr-SNOM was either limited to MHz repetition rate laser systems\cite{Eisele2014,Mrejen2019,Pizzuto2021}, or to using AFM tips with a lower resonance frequency\cite{Sternbach2017}.
Introducing phase-domain sampling, Wang et al.\cite{Wang2016} were able to circumvent the Nyquist limit in sampling frequency for a self-homodyne (shd) detection scheme.
This extends the possible implementation of SNOM to kHz laser systems, providing stable high peak powers useful e.g.\ for parametric amplification, white light generation, photo-switching, or nonlinear spectroscopy.
Phase-domain sampling was extended to pseudo-heterodyne modulation (pshet)\cite{Ocelic2006}, by sampling modulation phases independently for every laser shot using quadrature-assisted discrete (QUAD) demodulation\cite{Palato2022}.
However, tr-SNOM using phase-domain sampling has yet to be demonstrated.

In this paper we extend QUAD sampling to ultrafast pump-probe measurements.
Using a 200\,kHz laser system, we present the first realization tr-SNOM acquired in the phase-domain, completely independent of the sampling frequency.
First, we derive a tr-SNOM signal from basic principles that is independent of any physical model of the near-field interaction.
Our new experimental approach retrieves the static and excited SNOM signal simultaneously and independently after a cascaded demodulation scheme.
We are able to distinguish background and near-field contributions in the harmonic orders of the tr-SNOM signal, and confirm the spatial confinement in a \ce{WS2} monolayer/multilayer boundary region.
By isolating and quantifying the dynamic tip response, we model the near-field response of \ce{WS2} and obtain a dynamic dielectric function.

\section{A general time-resolved SNOM signal}
The essence of optical pump-probe, or any ultrafast measurement, is to compare a static signal (\emph{pump-off}) to a dynamic signal a certain time after excitation (\emph{pump-on}).
We now define a time-resolved SNOM signal, based on these two measurements.
In static SNOM, we measure the electric field scattered by the tip-sample junction $E_\mathrm{sca, off} = \sigma_\mathrm{off} E_0$, described by the scattering coefficient $\sigma$ and the incident light $E_0$.
The scattered light consists of optical near-field (NF) and a background signal (BG): $\sigma_\mathrm{off} = \sigma_\mathrm{NF} + \sigma_\mathrm{BG}$.
The experimental challenge in SNOM is to isolate $\sigma_\mathrm{NF}$.
Detecting visible wavelengths, this is efficiently achieved demodulating high harmonics of tapping modulation, and introducing a local oscillator $E_\mathrm{ref}$ for interferometric detection.
The photodiode measures the combined intensity of NF, BG, and reference signal (ref).
\begin{align}
    \vert E_\mathrm{sig,off}\vert^2 &= \vert E_\mathrm{NF} + E_\mathrm{BG} + E_\mathrm{ref}\vert^2\\
    &= \vert E_\mathrm{NF}\vert^2 + \vert E_\mathrm{BG}\vert^2 + \vert E_\mathrm{ref}\vert^2
    + E_\mathrm{NF}^{}E_\mathrm{BG}^\ast + E_\mathrm{ref}(E_\mathrm{NF}^\ast + E_\mathrm{BG}^\ast) + c.c.
    \label{eq:E_sig_off}
\end{align} 
where \ENF{} and \EBG{} are the near-field and background contributions to the detected signal.
Pshet modulation employs a periodic modulation of the reference phase, thereby allowing the isolation of $\sigma_\mathrm{off} \propto E_\mathrm{NF} + E_\mathrm{BG}$ in the signal spectrum\cite{Ocelic2006}.
Demodulation at harmonics of AFM tapping modulation then suppresses $E_\mathrm{BG}$\cite{Wurtz1999,Labardi2000}.
Without a local oscillator, the SNOM signal is dominated by the term $\ENF^{}\EBG^\ast$, containing a multiplicative background contribution in all harmonics\cite{Ocelic2006}.

Optical pump-probe measurements are a third-order nonlinear process\cite{Mukamel1999}, and as such an addition to the static signal.
The modified material response after photo-excitation by the pump pulse leads to additional time-dependent contributions in both NF and BG signals:  $\Delta E(t)=\Delta \ENF(t)+\Delta \EBG(t)$.
The reference field does not interact with the sample, and therefore is not modified by the pump.
The total signal at the detector is thus:
\begin{align}
    E_\mathrm{sig,on} &= E_\mathrm{sig,off} + \Delta E \\
    \vert E_\mathrm{sig,on}\vert^2 &= \vert E_\mathrm{sig,off}\vert^2 + \vert\Delta E_\mathrm{NF}\vert^2 + \vert\Delta E_\mathrm{BG}\vert^2 + \Delta E_\mathrm{NF}\left(E_\mathrm{NF}^\ast + E_\mathrm{BG}^\ast + \Delta E_\mathrm{BG}^\ast\right)\label{eq:E_sig_on}\\
    &+ \Delta E_\mathrm{BG}\left(E_\mathrm{NF}^\ast + E_\mathrm{BG}^\ast\right)\nonumber + E_\mathrm{ref}\left(\Delta E_\mathrm{NF}^\ast + \Delta E_\mathrm{BG}^\ast\right) + c.c.\nonumber
\end{align}
As in the static case, pshet modulation isolates the contribution $\sigoff+ \Delta\sigma(t)$.
Subtracting the static signal yields the pump-induced change in scattered light intensity.
\begin{align}
    \Delta\sigma(t) &= \sigma_\mathrm{on}(t) - \sigma_\mathrm{off}
    \label{eq:delta_sigma}
\end{align}

Using self-homodyne modulation (shd), $E_\mathrm{ref}=0$, and the dynamics of $\Delta E_\mathrm{NF}(t)$ are overshadowed by $\Delta E_\mathrm{BG}(t)$ in terms such as $\Delta E_\mathrm{BG}(t)E_\mathrm{NF}^\ast$, which is the interference between the spatially resolved static near-field and the time-dependent background.
The effect has initially been demonstrated by imaging on \ce{VO2} films\cite{Sternbach2017}, and is further demonstrated theoretically and experimentally by time-resolved measurements provided in the supplement.
The dominance of multiplicative background is completely analogous to the static case, where it was shown that even the higher order NF is modulated by the far-field background signal\cite{Ocelic2006}.
To isolate the time-resolved near-field signal, it is therefore necessary to employ a triple modulation scheme: tapping modulation, interferometric detection and pump modulation.
This enables the measurement of $\Delta\sigma(t)$ in amplitude and phase.

\begin{figure}
    \includegraphics{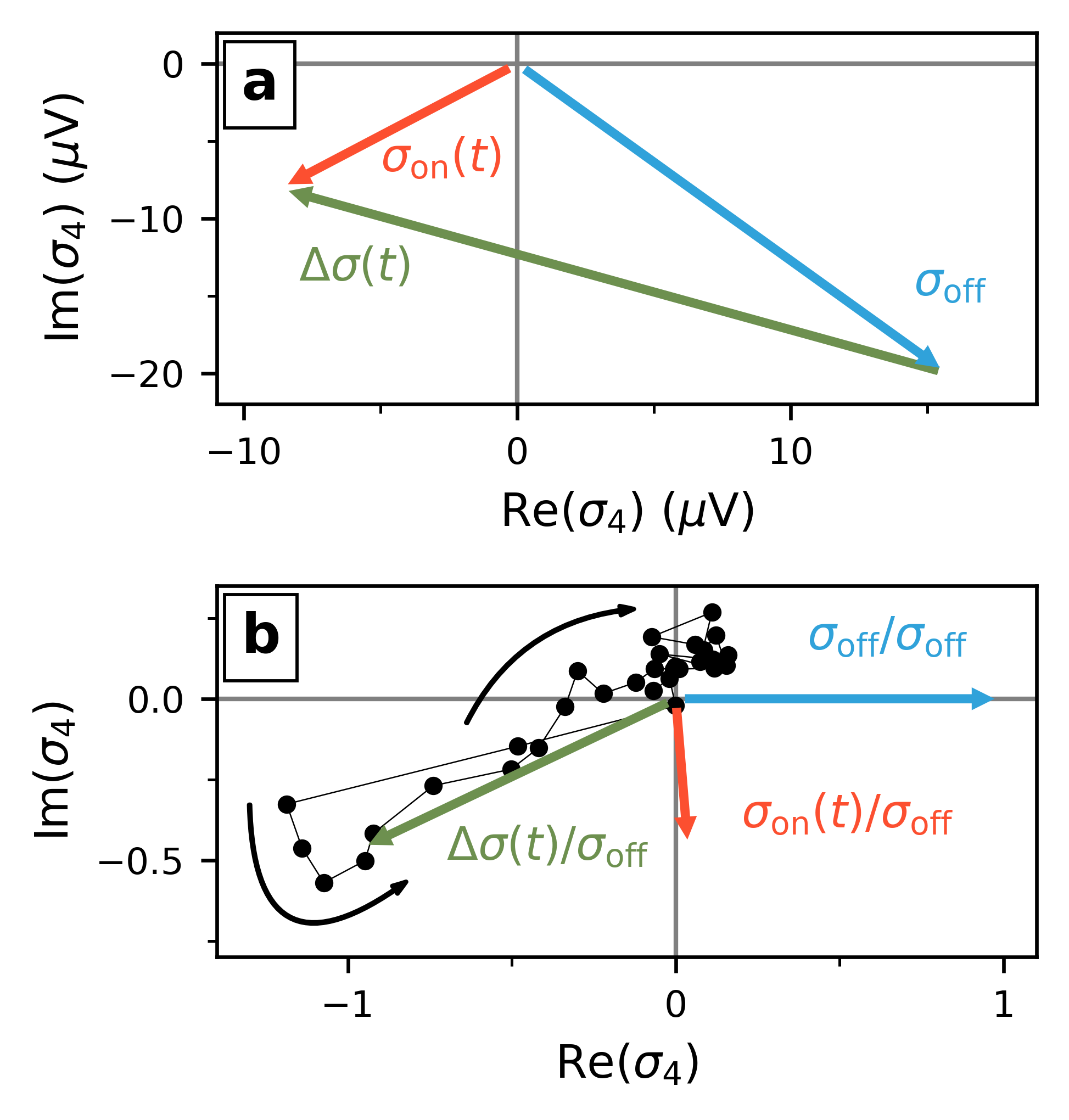}
    \caption{
        Visualization of the complex SNOM signal in a pump-probe measurement.
        (a): $\sigma_\mathrm{on}$ is the signal after pump excitation (\emph{pump-on}), $\sigma_\mathrm{off}$ is the SNOM signal without pump excitation (\emph{pump-off}), $\Delta\sigma$ is the pump-induced difference of the two.
        (b): All signals normalized by the probe-only signal $\sigma_\mathrm{off}$.
        Example experimental time-resolved SNOM signal shown in black.
        Arrows indicate the direction of time.
    }
    \label{fig:phasors}
\end{figure}

To illustrate pump-probe modulation in tr-SNOM and to help define signal normalization, we will be using a phasor analysis approach.
\textbf{\Cref{fig:phasors}}(a) visualizes \Cref{eq:delta_sigma}.
$\sigma_\mathrm{off}$ is the pshet signal acquired before photo-excitation.
It is demodulated at tapping harmonic $n=4$, suggesting dominance of $E_\mathrm{NF}$.
$\sigma_\mathrm{on}(t)$ is the pshet signal acquired after photo-excitation, and according to \Cref{eq:E_sig_on} is the isolation of $\Delta E_\mathrm{NF}$ in addition to the static signal.
The difference $\Delta\sigma(t)$ is then proportional to $\Delta E_\mathrm{NF}$.

Although isolating $\Delta\sigma_\mathrm{NF}(t)$ already yields the time-resolved near-field signal, its magnitude and phase are dependent on those of $\sigoff$, which are determined by experimental factors that are in general not known and might drift during the measurement.
In order to avoid these effects, and to recover the sample material response, we normalize the signal by $\sigoff$, as illustrated in \Cref{fig:phasors}(b).
This has the effect to rotate and scale all the phasors.
$\sigma_\mathrm{off}/\sigma_\mathrm{off}$ becomes unity by definition.
$\sigma_\mathrm{on}(t)/\sigma_\mathrm{off}$ describes the relative amplitude and the difference in phase of \emph{pump-on} and \emph{pump-off} signals.
The time-resolved SNOM signal is then:
\begin{align}\label{eq:delta_sigma/sigma}
    \frac{\Delta\sigma(t)}{\sigma_\mathrm{off}} = \frac{\sigma_\mathrm{on}(t) - \sigma_\mathrm{off}}{\sigma_\mathrm{off}} = \frac{\sigma_\mathrm{on}(t)}{\sigma_\mathrm{off}} - 1
\end{align}
as shown by the green arrow. 

This normalization has a few advantages. It does not depend on physical assumptions about the tip-sample interaction.
The ratio $\Delta\sigma(t)/\sigma_\mathrm{off}$ cancels arbitrary signal scaling and phase shifting due to experimental geometry.
This signal is therefore not distorted by experimental factors such as the intensity of the probe pulse, quality of the interferometric contrast, alignment to the AFM tip, magnitude of NF signal, or arbitrary optical phase offset, nor by their drifts.
The magnitude and phase of $\Delta\sigma(t)/\sigma$ can be interpreted directly.
For example, a phase of $\pi$ indicates a pump-induced reduction in signal amplitude, while a phase of $0$ corresponds to an increase.
$\Delta\sigma(t)/\sigma_\mathrm{off}$ also provides ways to assess the validity of the data: causality imposes $\Delta\sigma(t)/\sigma_\mathrm{off} = 0$ for $t<0$.
The signal must also go to 0 when the pump is blocked, and the magnitude of $\Delta\sigma(t)/\sigma_\mathrm{off}$ allows optimization of the pump-probe overlap.
The signal should also vanish scattering off of media that do not interact with the pump, such as fused silica in the visible.
Furthermore, the ratio $\sigon/\sigoff$ allows compensation of probe beam intensity fluctuations, thereby improving signal to noise ratio.

The black dots in \Cref{fig:phasors} show experimental measurements of $\Delta\sigma(t)/\sigma_\mathrm{off}$, with the direction of time indicated by black arrows.
At negative times, the signal is initially around the origin.
At $t=0$, it suddenly increases in amplitude and moves to the lower left quadrant.
Then, the pump-probe signal decays to a longer lived component with a phase $\sim\pi/2$, decaying to 0 as $\sigma_\mathrm{on}(t)$ approaches $\sigma_\mathrm{off}$ for large $t$.
This is the expected behavior for a pump-probe signal that obeys causality.
The tr-SNOM signal will be referred to as $\Delta\sigma/\sigma$ in the text below.

\section{Methods}

\begin{figure}
    \includegraphics{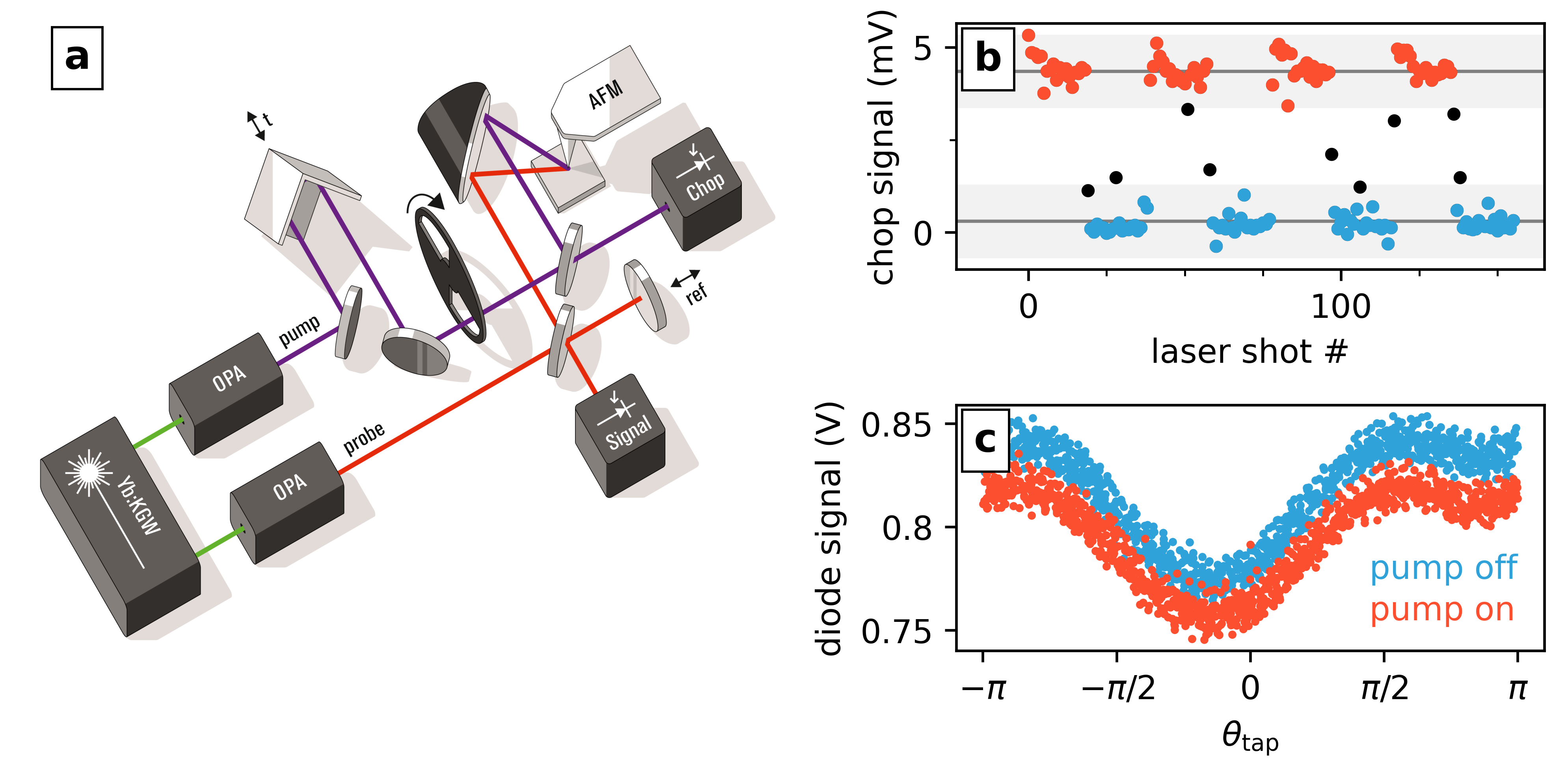}
    \caption{
        (a): Schematic of time-resolved SNOM optical setup.
        The pump beam path is delayed with respect to the probe beam path, and chopped using an optical chopper wheel. The state of the pump is detected using the "chop" photodiode.
        The probe beam path enters a Michelson interferometer for pseudo-heterodyne modulation. Graphic by Peter Schwendke. Reproduced with permission.
        (b): Photodiode voltage determining the state of the chopper wheel.
        Red: \emph{pump-on}, blue: \emph{pump-off}, black: pump pulses clipped by the chopper wheel are disregarded during analysis.
        (c): Comparison of the voltage of the photodiode after optical excitation (\emph{pump-on}) and without (\emph{pump-off}).
        The intensity of individual laser pulses shows tapping modulation with multiple harmonics, and a pump-induced change in signal level.
	}
    \label{fig:setup_signal}
\end{figure}

Our tr-SNOM setup consists of a commercial SNOM (neaSNOM, neaSpec) and a 200\,kHz femtosecond laser (PHAROS, Lightconversion) driving two independent optical parametric amplifiers (ORPHEUS, Lightconversion) for pump and probe pulses.
\Cref{fig:setup_signal}(a) shows a scheme of the optical setup.
The pump pulse enters a delay stage and is chopped using a chopping wheel, the status of which is detected with a photodiode for every laser shot.
This assignment enables the simultaneous acquisition of $\sigma_\mathrm{on}$ and $\sigma_\mathrm{off}$ during the integration time, reducing low frequency noise contributions.
The probe pulse is spatially filtered (not shown) and enters a Michelson interferometer with the sample in the signal arm and a vibrating mirror in the reference arm, for established pshet modulation.
Pump beam and signal arm are combined in a parabolic mirror and focused on an AFM tip operating in tapping mode.
The parabolic mirror reflects the light scattered at the tip-sample junction back into the interferometer.
Following the pshet modulation scheme described by Ocelic et al.\cite{Ocelic2006}, the interference of signal and reference arm is detected at the photodiode.

In order to measure the tr-SNOM signal, we extend the QUAD stroboscopic scheme as follows.\cite{Palato2022}.
The continuous modulation signals, i.e.\ deflection laser and piezo voltage of the reference arm, are filtered and used to generate two copies: one in-phase ($x$) and one with a $\pi/2$ phase shift ($y$).
$x$ and $y$ are effectively cosine and sine of the modulation phase, which can then be recovered as $\theta=\mathrm{arctan2}(y/x)$. This calculation is performed for both the tapping and reference arm modulation, yielding $\theta_\mathrm{tap}$ and $\theta_\mathrm{ref}$.
As in the static case, the two ($x,y$) signal pairs are detected using a data acquisition card, alongside the voltage at the signal photodiode.
For time-resolved SNOM, we add one detection channel: the status of the chopping photodiode.
The acquisition of all channels is synchronized using a 200\,kHz sampling clock obtained from the laser amplifier.
Experimental control, signal acquisition and analysis are accomplished using dedicated python scripts, which are fast enough for live usage\cite{SNOMpad_trSNOM}.

To acquire $\sigma_\mathrm{on}$ and $\sigma_\mathrm{off}$ simultaneously, detected tuples of the photodiode signal $\propto\vert E_\mathrm{sig}\vert^2$, tapping modulation phase $\theta_\mathrm{tap}$ and reference modulation phase $\theta_\mathrm{ref}$ are first separated based on the chopping state.
\textbf{\Cref{fig:setup_signal}(b)} shows the voltage of the chopping photodiode for consecutive laser shots.
The chop signal has two mean values marked by grey lines: near 0 when the pump pulse is blocked by the chopper, and $\sim4.5$\,mV when the pump pulse passes and reaches the AFM tip.
The two cases are selected as quantiles over a set of samples, acquired during the integration time.
The first analysis step in demodulating the time-resolved signal is separating \emph{pump-on} and \emph{pump-off} tuples and neglecting shots where the pump is clipped by the chopper wheel.

\Cref{fig:setup_signal}(c) shows the photodiode signal in the tapping phase domain, measured for individual laser shots.
The \emph{pump-off} signal shows only the detected probe pulses where the pump was blocked.
The \emph{pump-on} signal shows only the detected probe pulses where the pump was not blocked.
Demodulation of the separated data points gives access to $\sigma_\mathrm{off}$ and $\sigma_\mathrm{on}(t)$, respectively.

To recover the pshet modulated time-resolved signal, both subsets of the data undergo identical demodulation steps.
The demodulation algorithm is slightly improved with respect to our previous work\cite{Palato2022}.
First, all samples of photodiode signal are averaged in a binned 2D phase domain spanned by $\theta_\mathrm{tap}$ and $\theta_\mathrm{ref}$.
This creates an array of real values with tapping modulation along $\theta_\mathrm{tap}$ and pshet modulation along $\theta_\mathrm{ref}$.
Performing a fast Fourier transform along $\theta_\mathrm{tap}$ gives us access to the harmonics of tapping modulation.
\begin{align}
    \vert E_\mathrm{sig}\vert^2 &= \sum_n\tau_n(\theta_\mathrm{ref}) e^{in(\theta_\mathrm{tap} - \theta_C)}
    \label{eq:Esig_tapping_harmonics}
\end{align}
where $\tau_n(\theta_\mathrm{ref})$ now depends on the tapping demodulation order $n$ and the phase of the reference modulation $\theta_\mathrm{ref}$.
Describing the motion of the tip as a harmonic oscillation, all harmonics of the tapping modulation are symmetric around the phase of contact $\theta_C$.
Correcting that phase offset, the harmonic spectrum becomes real and the signal hermitian, i.e.\ $\tau_n$ becomes purely real.
The imaginary part of $\tau_n(\theta_\mathrm{ref})$, as returned by the fast Fourier transform, can therefore be disregarded as noise.

$\tau_n(\theta_\mathrm{ref})$ describes the pshet modulation, i.e.\ the interference of signal arm and reference arm at the photodiode, as contained in individual tapping harmonics $n$.
The reference arm undergoes an oscillating phase modulation with modulation phase offset $\theta_0$ and optical phase offset $\psi_R$.
\begin{align}
    E_\mathrm{ref} &= \rho\,e^{i\psi}E_0\label{eq:Eref}\\
    \psi &= \gamma \sin(\theta_\mathrm{ref} - \theta_0) + \psi_R
\end{align}
$\rho$ is a scaling factor and $\gamma$ the phase modulation depth in radians.
Comparing to \Cref{eq:E_sig_off,eq:E_sig_on}, we can separate the contributions of the SNOM signal according to their dependence on $\theta_\mathrm{ref}$.
\begin{align}
    \tau_{n}(\theta_\mathrm{ref}) &= c_n + E_\mathrm{ref}^{}(\theta_\mathrm{ref})\sigma_{n}E_0^\ast + c.c.\label{eq:tau_off}
\end{align}
We assume stability of the reference arm parameters $\rho$, $\gamma$, $\theta_0$, and $\psi_R$ during the acquisition time.
This enables us to fit $c_n + \mathrm{Re}(\rho e^{i\psi})$ to $\tau_n(\theta_\mathrm{ref})$ and attribute the fitting parameters $\rho$ and $\psi_R$ to amplitude and phase of $\sigma_n$ respectively.
$\gamma$ and $\theta_0$ do not depend on the harmonic order and can be fixed fitting $\tau_0$, to stabilize the fit for higher harmonics.
This procedure is performed on both \emph{pump-on} and \emph{pump-off} data points, generating $\sigma_{n,\mathrm{on}}$ and $\sigma_{n,\mathrm{off}}$.
The complex time-dependent SNOM signal $\Delta\sigma_n/\sigma_n$ is formed using \Cref{eq:delta_sigma/sigma}.
This demodulation algorithm provides an improved signal to noise ratio and experimental stability compared to our previous algorithm\cite{Palato2022}.
The next section will experimentally demonstrate this approach.

\section{Time-resolved near-field}
We now assess the spatial dependence and kinetics of the tr-SNOM signal $\Delta\sigma/\sigma$ to confirm our method is able to isolate the time-resolved NF contribution. 
A common method to prove NF character are signal-distance curves, also called retraction or approach curves\cite{Specht1992,Inouye1994,Hillenbrand2000,Raschke2003}, which we discuss first.
In our setup the AFM tip, as the main scatterer, stays roughly in the same position, while the sample is moved downward by the piezo stage.
The NF signal $\sigma_\mathrm{NF}$, acquired during this retraction, can be distinguished from the background $\sigma_\mathrm{BG}$ by its stronger dependence on the tip-sample distance.
As the tip-sample distance d$z$ is increased, amplitude and phase of $\sigma_\mathrm{NF}$ stay constant while the tip is still in contact.
The signal amplitude then decays within the first $\sim 10\,$nm after loosing tip-sample contact.
A longer decay, oscillations, shifting phase, or residual signal indicate a background character that is not necessarily localized at the tip-sample junction.
These background signals can be due either to scattering from the sample, or from the tip.

\begin{figure}
    \includegraphics{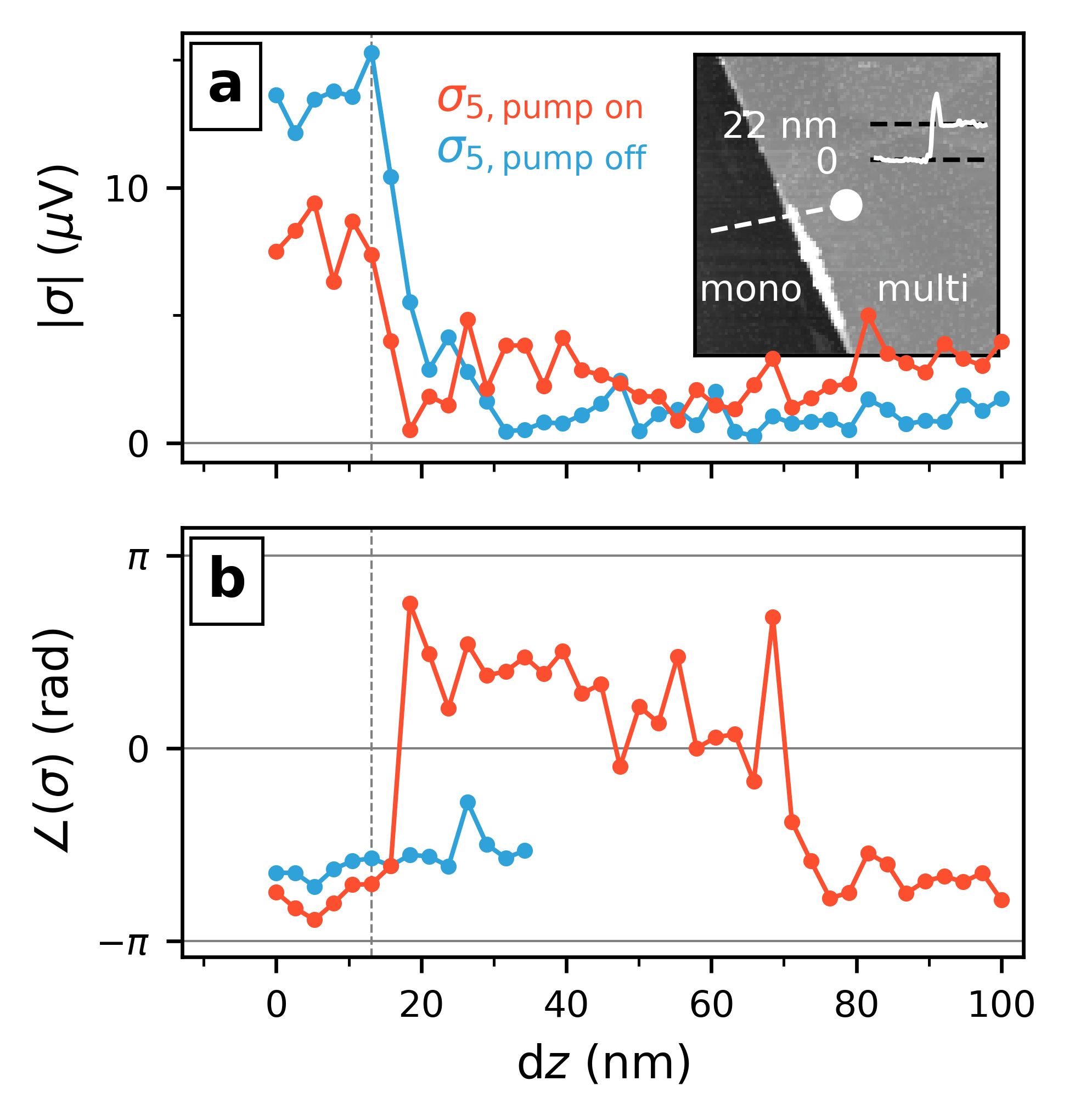}
    \caption{
        SNOM signal-distance curves taken at location marked in AFM inset.
        (a): Comparison of SNOM signal amplitude $\vert\sigma_5\vert$ vs tip-sample distance d$z$, taken before (\emph{pump-off}) and 0.5\,ps after (\emph{pump-on}) optical excitation.
        The dashed line marks the snap-off distance where the AFM tip looses contact to the sample, defined by the maximum AFM tapping amplitude during retraction.
        (b): Optical phase of $\sigma_5$ with and without optical excitation.
        At distances where the $\vert\sigma_5\vert$ drops below the noise floor, the phase becomes ill-defined and is not plotted (phase-blanking\cite{Trebino2000}).
        The inset shows an AFM image of the acquisition site with a height profile taken along the dashed line.
    }
    \label{fig:pumpprobe_retraction}
\end{figure}

\textbf{\Cref{fig:pumpprobe_retraction}} shows retraction curves taken on multilayer \ce{WS2} using an uncoated \ce{Si} tip.
\Cref{fig:pumpprobe_retraction}(a) shows the amplitude of NF signal $\sigma_5$ for both \emph{pump-on} and \emph{pump-off}, while (b) shows the phase.
The acquisition site is marked by a white dot in the AFM topography image of the multilayer region, shown in the inset in (a).
Signal amplitude and phase are mostly constant in contact.
After the AFM tip snaps off-contact (dotted line), the \emph{pump-off} signal (blue) shows a fast amplitude decay with a constant phase, suggesting NF character.
The \emph{pump-on} signal (red) shows a reduced amplitude in contact, also a rapid amplitude decay, but then a residual signal with a phase that shifts from $\pi/2$ to $-\pi$ with increasing tip-sample distance.
The residual signal is likely a dynamic response by the uncoated \ce{Si} tip, and will be discussed in \Cref{sect:tip_contribution}.
The retraction curves are corrected for a constant complex offset at large $\mathrm{d}z$, which we attribute to scattering off of the AFM tip without sample interaction.
Retraction curves using \ce{PtIr} coated tips, as commonly used in SNOM measurements, show similar results, but in general less pump-probe contrast and are presented in the supplement.
The decay of the SNOM signal within $\sim 10$\,nm implies its localization at the AFM tip, and the dominance of the NF contribution to \sigon{} while in contact.

\begin{figure}
    \includegraphics{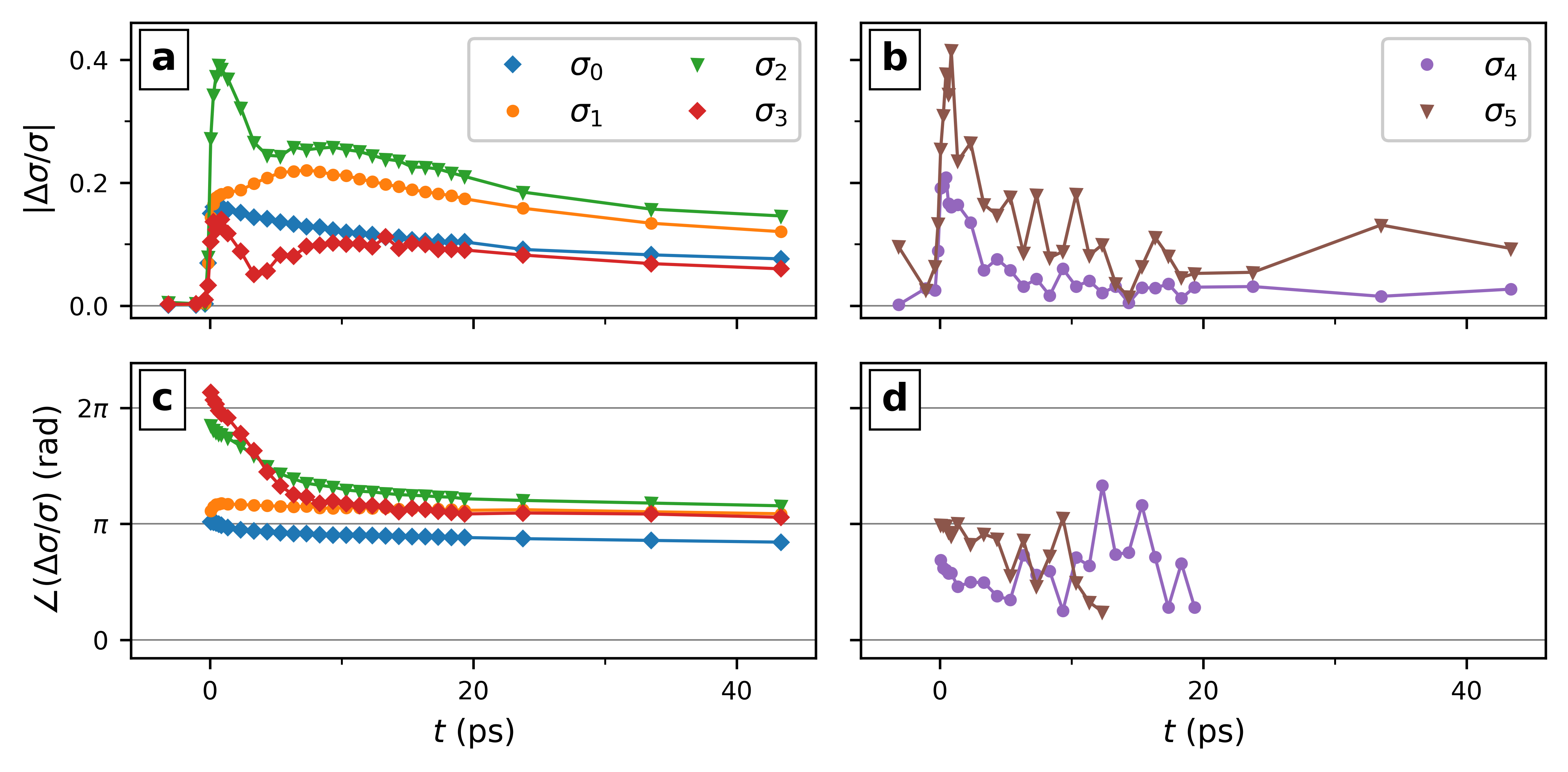}
    \caption{
        Time-resolved SNOM signal on multilayer \ce{WS2} for demodulation order $n=0$ -- 5.
        (a) and (b) show the signal amplitude vs delay time $t$, while (c) and (d) show the optical phase.
        $\sigma_0$ is the pure far-field contribution.
        $\sigma_{1-3}$ show varying contributions with distinct optical phase and varying amplitude.
        $\sigma_{4,5}$ show a dynamic signal decaying below the noise floor after $\sim10$\,ps, where it is phase-blanked.
        Location marked in inset in \Cref{fig:pumpprobe_retraction}(a).
	}
    \label{fig:multilayer_dynamics}
\end{figure}

To further confirm the identification as pump-modulated NF, \textbf{\Cref{fig:multilayer_dynamics}} compares  the time-resolved SNOM signal $\Delta\sigma/\sigma$ as a function of pump-probe delay time at various orders.
\Cref{fig:multilayer_dynamics}(a) and (c) show amplitude and phase of the lower harmonics, whereas (b) and (d) show the amplitude and phase of the higher harmonics.
$\sigma_0$ is pure far-field kinetics, not modulated by AFM tapping, and shows an abrupt rise at $t=0$, followed by a reduction in SNOM signal amplitude with phase $\sim\pi$.
The same contribution is visible in $\sigma_{1-3}$ at long times, but a new contribution with a phase close to 2$\pi$ and a faster time constant is visible in $\sigma_1$ and dominates the very early response in $\sigma_2$ and $\sigma_3$.
The kinetics of amplitude and phase appear non-trivial due to interference of the two complex-valued contributions, while being a simple addition in the complex plane, similar to the data points plotted in \Cref{fig:phasors}(b).
Finally, $\sigma_{4,5}$ in \Cref{fig:multilayer_dynamics}(b) and (d) show a contribution with a phase close to $\pi$.
The decay is faster than the far-field kinetics and drops beneath the noise floor after $\sim15$\,ps.
All harmonics $\Delta\sigma_n/\sigma_n$ show no signal at $t<0$ and rise abruptly at $t=0$, confirming the isolation of pump-induced changes.

The progressive change of signal contributions is in agreement with the suppression of diffraction-limited background by  higher-harmonic demodulation.
$\Delta\sigma_0/\sigma_0$ is dominated by $\sigma_{0,\mathrm{BG}}$ and $\Delta\sigma_{0,\mathrm{BG}}$, and shows pure BG kinetics evolving on long timescales.
$\Delta\sigma_{4,5}/\sigma_{4,5}$, on the other hand, is dominated by the NF contribution $\sigma_{4,5,\mathrm{NF}}$ and $\Delta\sigma_{4,5,\mathrm{NF}}$ showing distinct kinetics.
$\Delta\sigma_{1-3}/\sigma_{1-3}$ show a superposition, as expected.
In further analysis, $\sigma_4$ will be discussed instead of $\sigma_5$ as it is dominated by the same dynamic NF signal, but has an improved signal to noise ratio.

In contrast, tr-SNOM acquired without interferometric detection using self-homodyne (shd) modulation produces the same signal kinetics in all harmonic orders, as shown in Figure S1 in the supplement.
Without the use of a reference field $E_\mathrm{ref}$, the existence of any BG kinetics $\Delta E_\mathrm{BG}$ dominates all kinetic terms in \Cref{eq:E_sig_on}.
This way, it is not possible to distinguish any tapping harmonic from the far-field signal.
Even in the supposed case that all signal kinetics can be attributed to the NF, e.g.\ by experimental geometry, and setting the question of usefulness of tr-SNOM in this case aside, the lack of phase information provided by shd modulation makes an unambiguous identification of different contributions impossible.
The projection of a superposition of dynamic contributions, which are naturally complex valued, on a real value is not unique, prohibiting direct interpretation. 

\begin{figure}
    \includegraphics{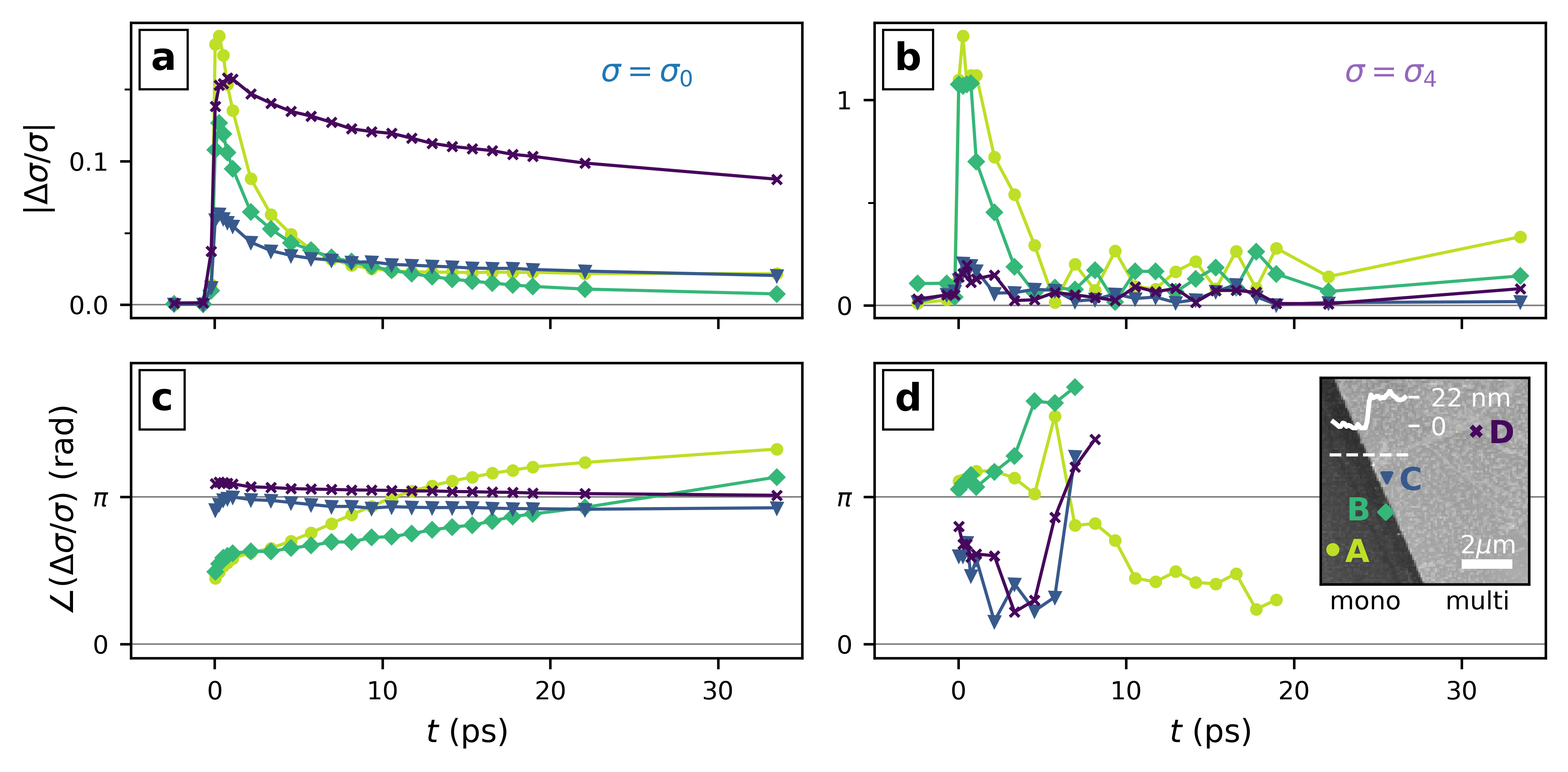}
    \caption{
        tr-SNOM signal at different locations on \ce{WS2} close to a monolayer-multilayer boundary, marked in AFM inset.
        (a) and (c) show amplitude and phase of the pure far field contribution $\sigma_0$.
        (b) and (d) show amplitude and phase of $\sigma_4$.
        While $\Delta\sigma_0/\sigma_0$ varies gradually moving across the boundary, $\Delta\sigma_4/\sigma_4$ shows distinct kinetics on monolayer and multilayer \ce{WS2}.
        The inset shows an AFM image of the acquisition site with a height profile taken along the dashed line.
	}
    \label{fig:spatial_dependence}
\end{figure}

To confirm the lateral localization of the dynamics observed in $\sigma_4$, we compare kinetics at closely spaced locations.
\textbf{\Cref{fig:spatial_dependence}} shows kinetic measurements taken at different positions across a boundary between \ce{WS2} monolayer and multilayer regions of the same sample (pos. A, B and C, D respectively).
\Cref{fig:spatial_dependence}(a) and (c) show the gradual change of the background signal $\sigma_0$, as the AFM tip moves across the boundary.
A contribution with phase $\sim\pi/2$ and faster time constant is more dominant on the monolayer, while a slower contribution with phase $\sim\pi$ can be observed on the multilayer.

$\sigma_4$, displayed in \Cref{fig:spatial_dependence}(b) and (d), features discrete kinetics on monolayer and multilayer regions, each showing similar and reproducible signal amplitude and phase.
Interestingly, pump-probe contrast of $\sigma_4$ is enhanced in the monolayer region, suggesting a complete bleach of the NF signal in the first few ps on the monolayer only.
In particular, the signals from pos. B and C, located very close to the junction, show a clear contrast.
The distinction, even close to the boundary, confirms the localization of the pump-probe signal at the AFM tip.

In summary, the isolation of the time-resolved NF contribution has been demonstrated in three ways.
\emph{Pump-on} and \emph{pump-off} signals in \Cref{fig:pumpprobe_retraction} show a clear pump modulation together with a strong tip-sample distance dependence, confirming vertical confinement.
The kinetics of different harmonic orders in \Cref{fig:multilayer_dynamics} reflect the description in \Cref{eq:E_sig_on}: the background dynamics are observed for $n=0$, mixed contributions can be identified for $n=1$ to 3, while $4^{th}$ and $5^{th}$ tapping harmonic can be attributed to NF kinetics.
The contributions each obey causality for $t<0$, and show distinct kinetics and optical phase.
We were also able to demonstrate high lateral resolution, by comparing the tr-SNOM signal on both sides of a material boundary in \Cref{fig:spatial_dependence}, showing confinement to the AFM tip.

\section{Discussion}
In the previous sections, we have shown that $\Delta\sigma_n/\sigma_n$ for sufficiently high $n$ is the time-resolved NF signal.
To present a physical interpretation, we will relate this tr-SNOM signal to the local dynamic dielectric function of the sample.
We model the tr-SNOM signal using the point dipole model, accounting for the time-dependent response of both the sample and tip.
We first measure the tip response using a reference measurement on \ce{SiO2}, a transparent dielectric.
Accounting for the tip response, we then analyze the response of monolayer and multilayer \ce{WS2}.

\subsection{Active tip dynamics}\label{sect:tip_contribution}

\begin{figure}
    \includegraphics{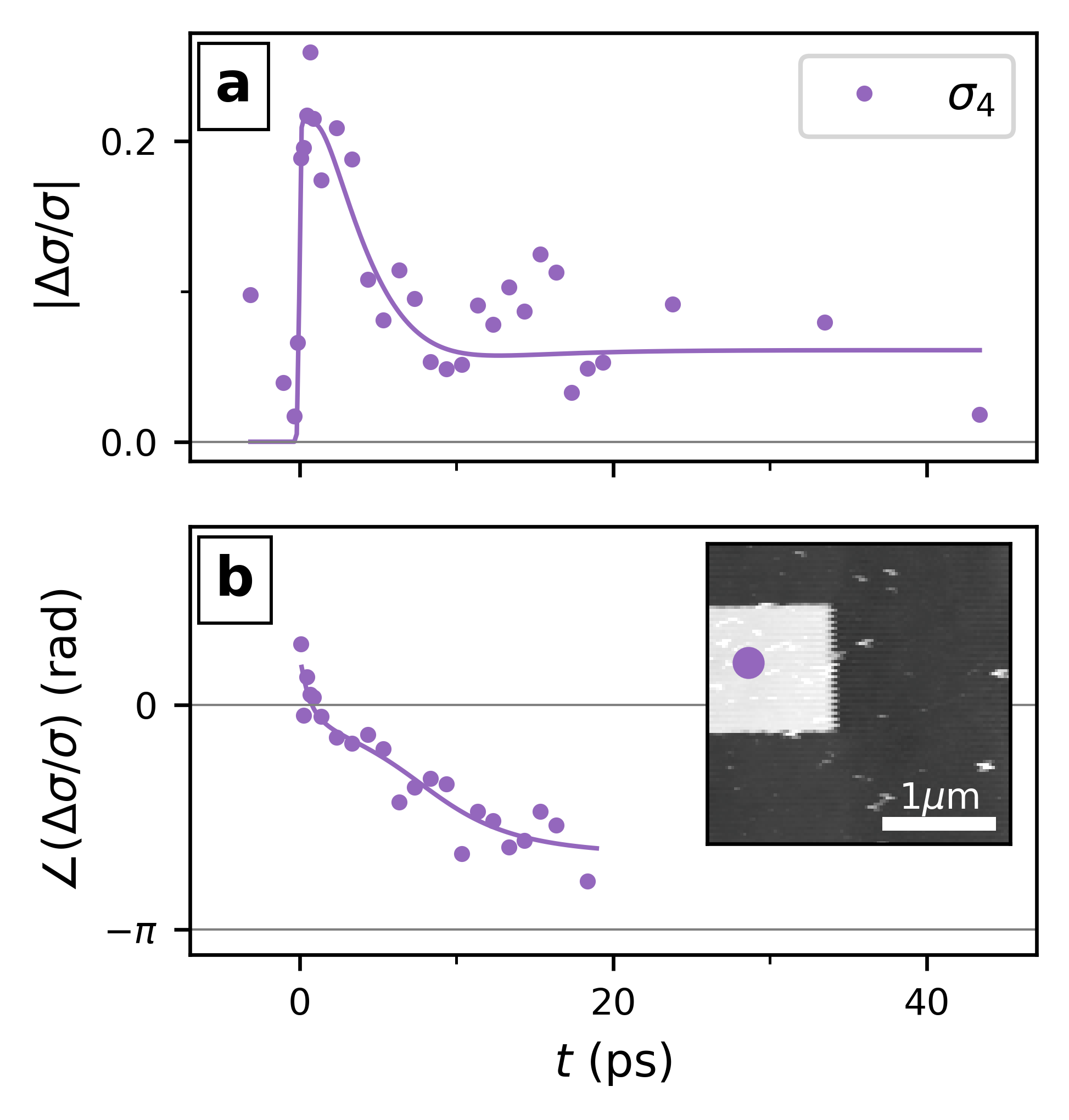}
    \caption{
        Time-resolved SNOM signal on \ce{SiO2}, acquired with an uncoated \ce{Si} tip.
        The fit, shown as a solid line, considers a time-dependent dielectric function of the tip, according to \Cref{eq:eps_dynamic}.
        The AFM inset marks the acquisition site with a dot, in an \ce{SiO2} area on an AFM test grating(TGQ1, Tips-Nano).
        The scan direction is alternating on the fast axis.
    }
    \label{fig:tip_contribution}
\end{figure}

\begin{table}
\begin{tabular}{@{}lccc@{}}
\toprule
Material & $A$ & $\tau$ (ps) & $\varepsilon_\mathrm{offset}$\\
\midrule
\ce{Si} tip & $-10.6 \pm 1.8$ & $4.5 \pm 3.2$ & $(0.9 + i2.0) \pm (2.1 + i0.7)$\\
\ce{WS2} monolayer & $(5.5 + i 4.0) \pm (0.8 + i 0.8)$ & $2.1 \pm 0.5$ & $(-0.1 - i 0.2) \pm (0.1 + i 0.1)$\\
\ce{WS2} multilayer & $(0.08 + i 0.01) \pm (0.03 + i 0.03)$ & $3.4 \pm 0.6$ & $(0.08 + i 0.01) \pm (0.03 + i 0.03)$\\
\bottomrule
\end{tabular}
\caption{
    Table of fit parameters determining the dynamic dielectric functions of \ce{Si} tip and \ce{WS2} sample, according to \Cref{eq:eps_dynamic}.
    The fits are shown in \Cref{fig:tip_contribution,fig:dynamics_fits}.
    The dynamic dielectric functions are plotted in \Cref{fig:dielectric_function}.
}
\label{tab:fit_results}
\end{table}

As mentioned above, uncoated \ce{Si} AFM tips show an increased pump-probe modulation compared to tips with a \ce{PtIr} coating.
To assess this enhancement, we measure the tr-SNOM signal on a \ce{SiO2} region of an AFM test grating using the same photon energies.
\textbf{\Cref{fig:tip_contribution}} shows amplitude and phase of the tr-SNOM signal acquired in the location marked in the AFM inset.
Since \ce{SiO2} is not excitable at the pump energy of 3.0\,eV, no contribution to a pump-probe signal is expected, and signal dynamics are instead attributed to the AFM tip.
The pump energy coincides with band nesting in the \ce{Si} band structure, exciting electrons efficiently into the conduction band.
The extra charge carriers lead to a dynamic contribution to the dielectric function of \ce{Si}\cite{Green2008,Combescot1985}.
We model the SNOM signal in \Cref{fig:tip_contribution} with a single exponential in the dielectric function of the tip with a complex offset to account for longer lived dynamics.
\begin{align}
    \varepsilon(t) &= \varepsilon_\mathrm{static} + \theta(t)\left(A\,e^{-t/\tau} + \varepsilon_\mathrm{offset}\right),\label{eq:eps_dynamic}
\end{align}
where $\theta(t)$ is the Heaviside step function, $\varepsilon_\mathrm{static,tip}=14.91+i0.12$ is the dielectric function of Si at 643\,nm, and the parameters $A_\mathrm{tip}$, $\tau_\mathrm{tip}$ and $\varepsilon_\mathrm{offset,tip}$ describe the dynamic response of the tip.
The modified dielectric function of the tip is used in a point dipole model\cite{Knoll2000}, to avoid additional assumptions on spatial distribution and homogeneity of the excited charge carriers inside the tip.
However, using the finite dipole model\cite{Cvitkovic2007} produces a similar result.
The static dielectric function of \ce{SiO2} is taken to be $\varepsilon_\mathrm{SiO_2}=2.12$\cite{Malitson1965}.
Fitting the time-resolved data, shown as a solid line, determines the dynamic dielectric function of the tip, with the fit parameters collected in \textbf{\Cref{tab:fit_results}}.
The time constant was determined as $\tau_\mathrm{tip} =(4.5 \pm 3.2)$\,ps, which is compatible with transient photoconductivity measured in \ce{Si} nanostructures.
Bergren et al. determined an initial decay of photoconductivity of $\sim2$\,ps after pumping at 3\,eV, attributed to cooling of hot carriers.
A subsequent decay on ns timescales is ascribed to long-lived excitons, both dynamics being dependent on the size of the nanostructures\cite{Bergren2014}.
Tr-SNOM measurement on \ce{Si} showed similar kinetics with a slightly increased modulation in $\vert\Delta\sigma/\sigma\vert$.
Parametrizing the dynamic dielectric function of the \ce{Si} tip as an initial pump-induced excitation, we are now able to disentangle tip and sample contribution to the tr-SNOM signal.

\subsection{Monolayer and multilayer dynamics}

\begin{figure}
    \includegraphics{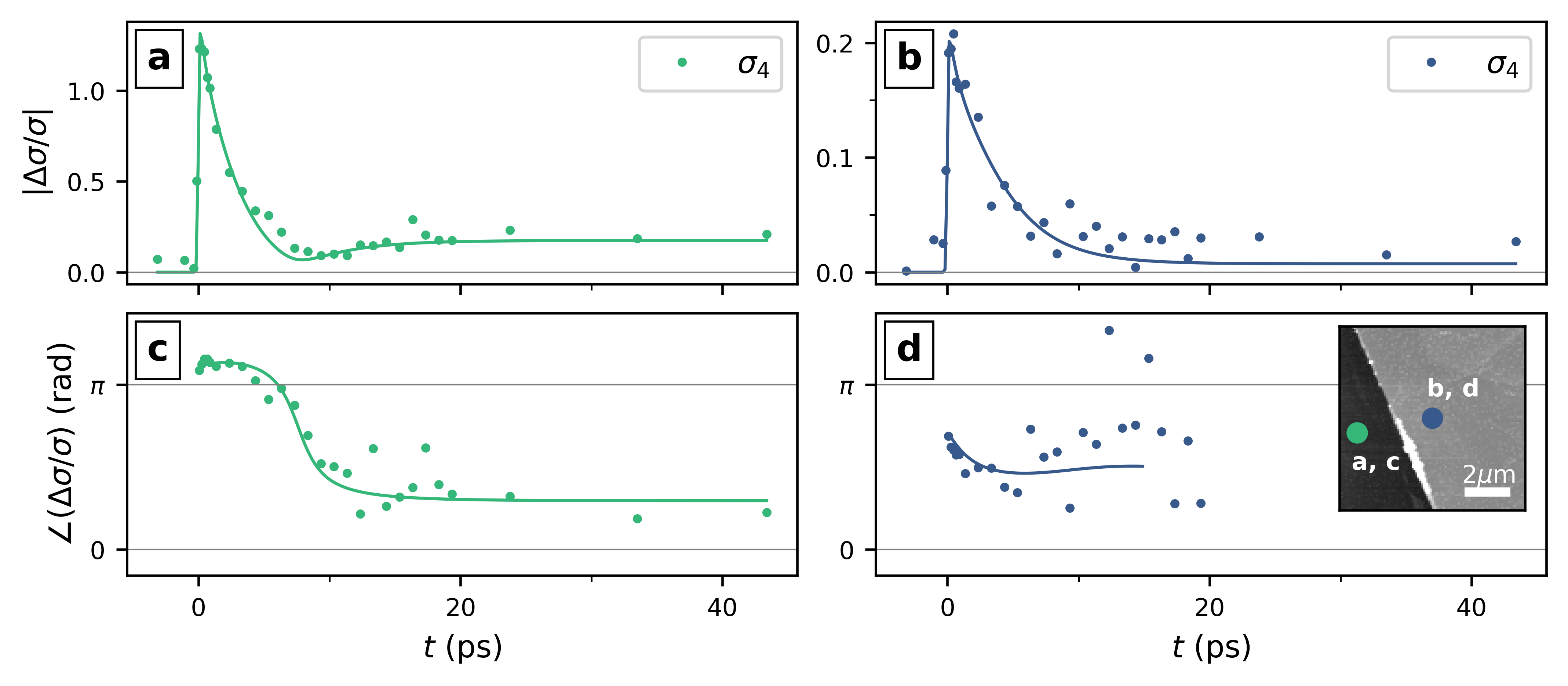}
    \caption{
        tr-SNOM signal at different locations on \ce{WS2} close to a monolayer-multilayer boundary, marked in AFM inset.
        (a) and (c) show amplitude and phase of near-field kinetics on monolayer \ce{WS2}.
        (b) and (d) show different kinetic contributions on the multilayer side of the boundary.
        The near-field was fitted, considering a time-dependent dielectric function of the \ce{Si} AFM tip, and the \ce{WS2} sample (\Cref{eq:eps_dynamic}).
	}
    \label{fig:dynamics_fits}
\end{figure}

\begin{figure}
    \includegraphics{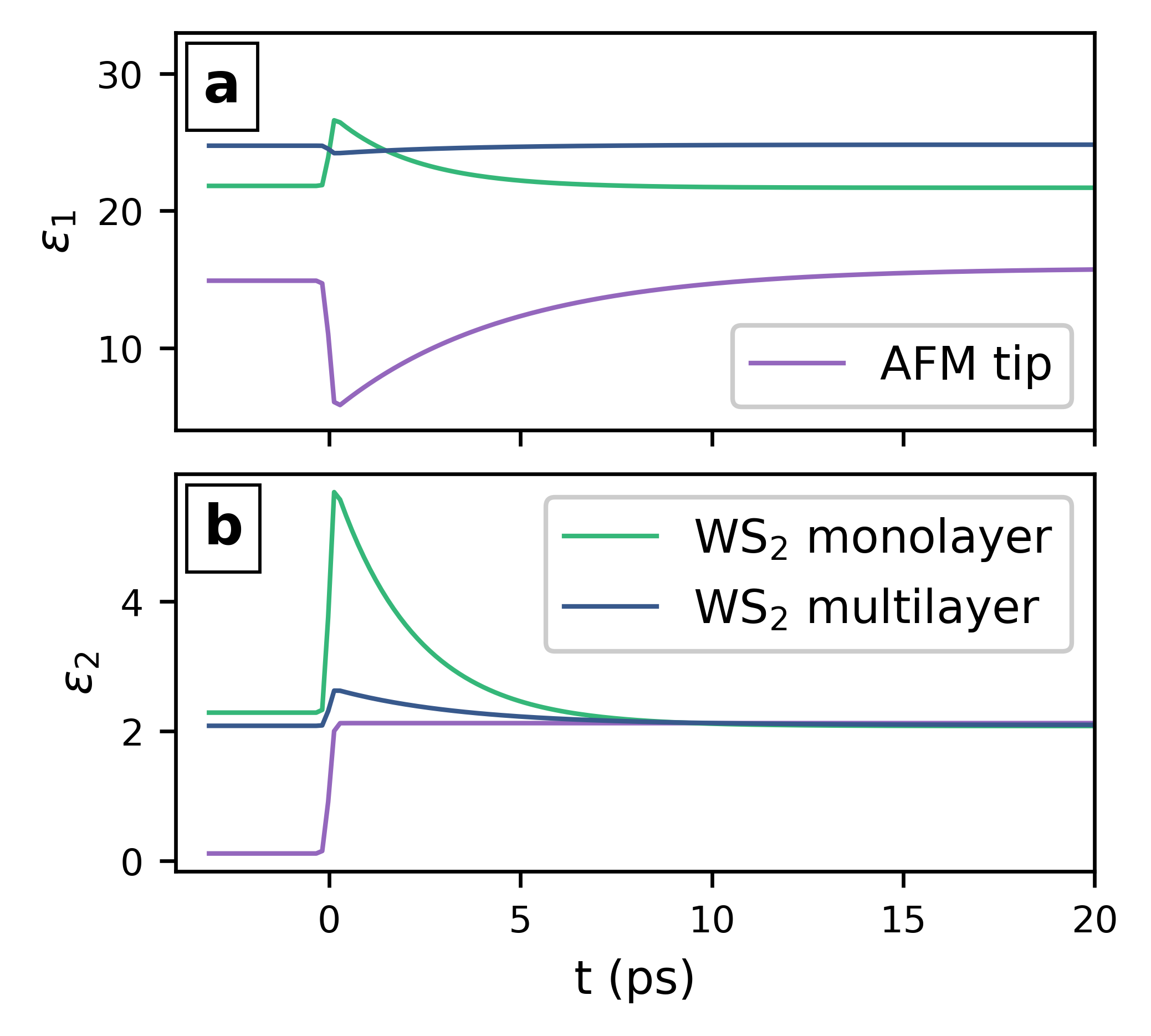}
    \caption{
        Modeled transient dielectric function of AFM tip (black), \ce{WS2} monolayer (blue), and \ce{WS2} multilayer (orange) after photo-excitation at $t=0$.
        Real part $\varepsilon_1$ in (a), and imaginary part $\varepsilon_2$ in (b). See text for details of the models.
	}
    \label{fig:dielectric_function}
\end{figure}

Using a similar model for the dielectric function of \ce{WS2}, the tr-SNOM signal provides a quantitative description of the ultrafast evolution of material properties at the nanoscale. 
\textbf{\Cref{fig:dynamics_fits}} shows NF kinetics ($\sigma_4$), taken on a \ce{WS2} monolayer region ((a) and (c)) and a multilayer region ((b) and (d)).
The kinetic traces are similar to the ones depicted in \Cref{fig:spatial_dependence}(b) and (d).
The data set plotted in \Cref{fig:dynamics_fits}(a) and (c) is also plotted in the complex plane as black dots in \Cref{fig:phasors}(b).
When modeling the tr-SNOM signal, we use the transient dielectric function of \ce{WS2} as described by \Cref{eq:eps_dynamic}.
$\varepsilon_\mathrm{static}$ at the probe energy of 1.9\,eV, just below the A-exciton resonance of 1.973\,eV, is taken from literature\cite{Calati2021}.
The transient dielectric function of the AFM tip, as determined in the previous section, was fixed while calculating the SNOM signal with the point dipole model.
The resulting fits are shown as solid lines in \Cref{fig:dynamics_fits}, describing both monolayer and multilayer dynamics.
The model provides a complex amplitude and offset, and a decay rate of a dynamic contribution.
\textbf{\Cref{fig:dielectric_function}} shows the evolution of the local dielectric functions, together with the previously determined dielectric function of the tip.
Both \ce{WS2} monolayer and multilayer show a pronounced photoinduced increase in $\varepsilon_2$.
While the monolayer also exhibits a distinct increase in $\varepsilon_1$, it is reduced on the multilayer.
Both measurements show time constants on the order of few ps, with $\tau_\mathrm{monolayer}=(2.1 \pm 0.5)$\,ps and $\tau_\mathrm{multilayer}=(3.4 \pm 0.6)$\,ps.
All fit parameters are collected in \Cref{tab:fit_results}.


The dynamics are in agreement with femtosecond spectroscopy results on these materials\cite{TandaBonkano2023,Villegas2024,Palummo2015,Calati2023}, which have found the dynamics can arise from a combination of peak shifts and amplitude reduction. These two phenomena cannot be disentangled from the current single-color measurements. To avoid over-determination of the model, we chose the phenomenological but general approach to directly model a dynamic dielectric function. Our results therefore demonstrate the isolation of different time-resolved and localized contributions to the tr-SNOM signal.
By first isolating the tip response, we were able to determine the time-resolved dielectric function of the sample.

\section{Conclusion}
We have defined a general tr-SNOM signal $\Delta\sigma/\sigma$ describing the time-resolved NF response of the sample, and presented an acquisition scheme using stroboscopic phase-domain sampling.
We demonstrated this method on \ce{WS2} monolayer and multilayer, using a 200\,kHz repetition rate laser to pump above bandgap at 3.0\,eV and probe the A-exciton response at 1.9\,eV.
The NF character was confirmed via signal-distance curves, by identifying BG and NF contributions in signal kinetics, and by resolving the localization of the dynamic NF at different positions in the monolayer-multilayer boundary region.
We quantified the dynamic response of the AFM tip by measuring the tr-SNOM signal on the transparent dielectric \ce{SiO2}.
Including the tip response in modeling the dynamic NF on \ce{WS2}, we were able to describe the local time-dependent dielectric function after photo-excitation of monolayer and multilayer.
We observed greatly enhanced pump-modulation of the monolayer signal compared to the multilayer, with a faster decay.
This marks the first realization of tr-SNOM using sparse sampling in the time-domain.
The tr-SNOM signal $\Delta\sigma/\sigma$ is material specific, localized on the nanoscale, and able to resolve ultrafast processes.

\section{Experimental Section}
In the experiments, the pump photon energy is tuned to 3.0\,eV (409\,nm) at $\sim1\,$nJ per pulse at the AFM tip, while the probe energy is 1.9\,eV (643\,nm) with $\sim0.125\,$nJ per pulse. The optical chopper wheel operates at $\sim 5$\,kHz.
The instrument response function has a width of $\sigma = (77 \pm 1)\,$fs, determined by fitting the onset of the far-field pump-probe signal with an error function.
The vibrating mirror modulates the optical path length of the reference arm at $\sim300$\,Hz.
The AFM tips are commercially available Si tips (Arrow NC, Nanoworld) or coated Si tips (Arrow NCPt, Nanoworld), with nominal resonance frequency of $\sim 275\,$kHz with an amplitude of 58\,nm in contact.

For alignment and characterization of the AFM tip, we use a calibration grating (TGQ1, Tips-Nano), featuring \ce{SiO2} squares on a \ce{Si} substrate.
Measurements on \ce{WS2} were performed on monolayer and multilayer regions on a \ce{Si} substrate terminated by $\sim300\,$nm of \ce{SiO2}, prepared by exfoliation\cite{Desai2016}.

\medskip
\textbf{Supporting Information} \par 
Supporting Information is available from the Wiley Online Library.

\medskip
\textbf{Acknowledgements} \par 
This work was funded by the Deutsche Forschungsgemeinschaft (DFG, German Research Foundation), Project-ID 182087777, SFB 951.
S.P. acknowledges funding from the Alexander von Humboldt foundation.
P.S. acknowledges support by the International Max Planck Research School for Elementary Processes in Physical Chemistry. 

\textbf{Conflicts of Interest} \par 
J.S., and S.P. are named inventors on EP 4 206 688 B1 and related patents (WO 2023/126233 A1; US 2025/0067770 A1), assigned to Humboldt-Universität zu Berlin and Max-Planck-Gesellschaft zur Förderung der Wissenschaften e.V., on which the present work builds. This patent family is licensed to attocube systems GmbH.

\medskip

%


\end{document}